\documentclass{emulateapj}
\usepackage{hyperref}
\usepackage{natbib}
\usepackage{graphicx}
\newcommand{\Msun}{\mathrm{M}_{\odot}}
\newcommand{\gcc}{\mathrm{g\ cm^{-3}}}
\begin{document}
\title{Magnetic Fields in Population III Star Formation}
\author{Matthew J. Turk\altaffilmark{1,2}, Jeffrey S. Oishi\altaffilmark{3,4},
Tom Abel\altaffilmark{3,4}, Greg L.~Bryan\altaffilmark{1}}
\email{matthewturk@gmail.com}
\altaffiltext{1}{Columbia University} 
\altaffiltext{2}{NSF CI TraCS Postdoctoral Fellow} 
\altaffiltext{3}{Stanford University} 
\altaffiltext{4}{SLAC National Accelerator Laboratory} 

\begin{abstract}
We study the buildup of magnetic fields during the formation of Population III
star-forming regions, by conducting cosmological simulations from realistic
initial conditions and varying the Jeans resolution.  To investigate this in
detail, we start simulations from identical initial conditions, mandating 16,
32 and 64 zones per Jeans length, and studied the variation in their magnetic
field amplification.   We find that, while compression results in some
amplification, turbulent velocity fluctuations driven by the collapse can
further amplify an initially weak seed field via dynamo action, provided there
is sufficient numerical resolution to capture vortical motions (we find this
requirement to be 64 zones per Jeans length, slightly larger than, but
consistent with previous work run with more idealized collapse scenarios).  We
explore saturation of amplification of the magnetic field, which could
potentially become dynamically important in subsequent, fully-resolved
calculations.  We have also identified a relatively surprising phenomena that
is purely hydrodynamic: the higher-resolved simulations possess substantially
different characteristics, including higher infall-velocity, increased
temperatures inside 1000 AU, and decreased molecular hydrogen content in the
innermost region.  Furthermore, we find that disk formation is suppressed in
higher-resolution calculations, at least at the times that we can follow the
calculation.  We discuss the effect this may have on the buildup of disks over
the accretion history of the first clump to form as well as the potential for
gravitational instabilities to develop and induce fragmentation.
\end{abstract}

\keywords{cosmology: theory --- galaxies: formation --- galaxies: HII regions
--- stars: formation}

\section{Introduction}
The formation of the first stars in the Universe is a process well-suited to
numerical computation.  While direct observations of these stars are still an
optimistic, yet-unrealized prospect \citep{2005ApJ...629..615W,
2009ApJ...698L..68F, 2005Natur.434..871F, 2011Natur.477...67C}, simulations are
able to begin with well-posed initial conditions and directly simulate their
formation.  These calculations, typically spanning many orders of magnitude in
both spatial and density scales \citep{2008AIPC..990...16T,
2008Sci...321..669Y}, include the effects of dark matter, hydrodynamics,
radiative cooling, chemical heating and cooling, and the non-equilibrium,
self-consistent evolution of multiple ionization and molecular states of the
gas \citep{abel97, anninos97, RA04, 2008AIPC..990...25G, 2008MNRAS.388.1627G}.

For nearly a decade, the consensus viewpoint had been that the first stars
formed in isolation, between $30-300~\Msun$ in mass, and only one per
$\sim 10^{6}~\Msun$ halo.  This had been confirmed by divergent methods: both
Smoothed Particle Hydrodynamics (SPH) simulations and Adaptive Mesh Refinement
(AMR) calculations showed similar results \citep{ABN02, 2003ApJ...592..645Y,
oshea07a}.  However, recently aspects of this picture are being challenged.
Advancements in the current generation of simulations have taken place in three
primary areas.  The first is that the effect of heating from the formation of
molecular hydrogen via three-body reactions \citep{2011ApJ...726...55T} is now
being taken into account; this will heat the gas during later stages of
collapse and through thermal regulation of the accretion rate produce higher
infall velocities onto the central core.  The second aspect is related to the
so-called ``Courant myopia'' of calculations proceeding to high densities.
Because the Courant timescale becomes extremely short at high densities, out of
necessity previous simulations stopped after the formation of the first
high-density core.  Recent simulations have followed accretion for up to
several thousand years, using the numerical technique of sink particles,
accreting Lagrangian particles representing protostars below the resolution
limit of the simulation \citep{2011ApJ...737...75G, 2011Sci...331.1040C,
2010MNRAS.403...45S}.  While this approach allows to follow the calculations
further in time it does lack the mathematical rigor that may allow one to prove
the result to be correct.  The final effect now being added is simply that of
sampling: in the published literature, very few calculations had been
performed; compared to the wealth of calculations of galaxy mergers, cluster
formation, and so on, the sampling of Population III star formation was
dramatically underserved, with only a handful of papers discussing even
multiple simulations \citep[see, e.g.,][]{oshea07a}.

These recent shifts have suggested that these halos have fragmented into either
a small-number multiplicity of protostars \citep{2009Sci...325..601T,
2010MNRAS.403...45S} or many small-mass pre-stellar objects
\citep{2011Sci...331.1040C, 2011ApJ...737...75G}.  
However, none of these simulations have included the effects of magnetic fields
in their calculations, nor have they self-consistently followed the growth of
seed magnetic fields over the collapse and virialization of these first
minihalos. Similarly, the fragmentation has been only studied for
times of less than 1\% of the accretion time scale of the massive
stars being studied. It has been difficult to show whether possible
early fragments would survive and not grow substantially in mass 
and simply merge with the central proto-star. 

The influence of magnetic fields on the collapse of the first stars has
received a renewed interest in recent years \citep{2008ApJ...685..690M}. Early
numerical simulations \citep{2008ApJ...688L..57X} including the Biermann
battery effect suggested that the dynamical effect of magnetic fields on
primordial gas is very small.  Though their resolution was rather low, these
simulations found that fields were primarily built up by collapse, leading to
$B \propto \rho^{2/3}$.  The thermodynamics of primordial gas are quite
sensitive to the $\mathbf{B} - \rho$ relationship via heating from ambipolar
diffusion \citep{2010ApJ...721..615S}.  Although magnetic fields are believed
to be unlikely to affect the characteristic fragmentation scale for Population
III stars, they may increase the temperature in very high density gas, leading
to higher accretion rates onto the protostars \citep{2009ApJ...703.1096S}. The
ability of turbulent fluid motions to drive dynamo action during primordial
protostellar collapse has been predicted analytically by
\citet{2010A&A...522A.115S} and confirmed numerically by
\citet{2010ApJ...721L.134S} and \citet{2011ApJ...731...62F} for idealized
collapse calculations. The latter work also suggested that a minimum resolution
of 32 elements per Jeans length is required to capture dynamo action, giving a
very strong motivation for higher resolution cosmological MHD calculations.
Properly resolved, dynamo amplification leads to exponential growth of the
field over a turbulent eddy turnover time, and can radically change the
strength of the magnetic field at any given density during collapse.  The
\citet{2011ApJ...731...62F} simulations considered the collapse of a
nearly-isothermal Bonnor-Ebert sphere with turbulence seeded by velocity
perturbations. While these conditions were idealized from earlier hydrodynamic
cosmological computations, we seek a more complete picture by following the
full magnetohydrodynamic (MHD) evolution of primordial gas from cosmological
initial conditions. This allows us to self-consistently study the interaction
of primordial gas dynamics, turbulence, and dynamo action.

In this paper, we present the first highly-resolved calculations of the formation
of the first stars in the Universe from cosmological initial conditions, taking
into account a full suite of chemical reaction rates, chemical heating due to
three-body reactions, magnetic fields from primordial seed fields, and a
resolution of 64 zones per Jeans length.  Furthermore, it has been conducted
with an open source, community-built simulation code available for inspection
and contribution.  In this paper, we examine both the large-scale and
small-scale amplification of the magnetic field, as well as the manner in which
resolution affects the chemo-kinetic state of the inner molecular cloud.  A
forthcoming paper will study the growth of turbulence in more detail.

\section{Methods and Simulations}

The simulations described in this work were conducted with the Enzo simulation
code\footnote{\url{http://enzo-project.org/}, described in
\cite{Bryan97,oshea04}, conducted with changeset \texttt{f3cf4f13e195}}.  Enzo
is an adaptive mesh refinement (AMR) simulation code, wherein the baryonic
fluid is discretized onto an Eulerian grid.  In cells where certain criteria
are violated (such as overdensity, Jeans parameter, slopes of fluid quantities)
higher resolution regions are inserted.  This process is applied dynamically
throughout the course of the simulation, ensuring adequate resolution of
physical processes on all length scales -- while the innermost regions of the
calculation may evolve more quickly dynamically, the entire simulation is
coupled and large-scale torques, inflow, and halo properties are
self-consistently included.

They were initialized with cosmological initial conditions generated from
Gaussian random noise, in a manner identical to that conducted in
\citet{ABN02,oshea04,2009Sci...325..601T,2010ApJ...725L.140T}.  The parameters
for generating the initial perturbations, as well as the partitioning of the
matter budget into dark and baryonic components, have been taken from WMAP7
results \citep{2011ApJS..192...14J}.  The simulations were initialized from a
single random seed at a redshift of 99, with their simulation volume centered
on the first massive halo to form in the simulation box,
$8.5\times10^{5}~\Msun$ at $z=19.72$.  For our initial conditions, we create a
set of nested grids initialized from the primordial power spectrum, which are
then allowed to dynamically refine as described below.  The initial structure
of our simulation was an outermost $128^3$, $300~\mathrm{kpc}~\mathrm{h}^{-1}$
(comoving) box, within which three subsequent, nested levels of refinement were
initialized, for a final dark matter mass resolution of $2.3 \Msun$ and a
baryonic cell width of $290~\mathrm{pc}~\mathrm{h}^{-1}$ (comoving).  The
highest-resolution region at time of initialization is a cube of side length of
$75~\mathrm{kpc}~\mathrm{h}^{-1}$ (comoving) and we allow further refinement
during the course of the simulation within a cube of side length
$12~\mathrm{kpc}~\mathrm{h}^{-1}$ (comoving) centered on the most massive halo
in the simulation at a redshift of $20$.

To solve the cosmological magnetohydrodynamic equations, we use the HLL Riemann
solver with piecewise linear reconstruction. The $\mathbf{\nabla \cdot B} = 0$
constraint is enforced using the Dedner scheme \citep[see][for a full
description]{2008ApJS..176..467W, 2010NewA...15..581W}.  We have utilized a
lower-order chemical solver than previous studies; in both
\cite{2010ApJ...725L.140T, 2009Sci...325..601T} a solver based on the work of
\cite{Verwer94gauss-seideliteration} was used.  Here we have used a modified
version of the solver from \cite{anninos97}, in the interest of quantifying the
influence of the magnetic field on the overall collapse, as well as the growth
of the magnetic field.  This choice of chemistry solver results in less
finely-balanced equilibrium between molecular hydrogen formation and
dissociation.  Despite that, as discussed below, there are substantial
differences in the molecular hydrogen fraction in these calculations, as well
as the temperature, as a function of resolution.  We have included the
formation heating from molecular hydrogen ($4.48~\mathrm{eV}$), and have used
the \cite{2008AIPC..990...25G} three-body formation rate.   Recent works
\citep{2011arXiv1108.5176N,2011ApJ...736..147G} have suggested that relative
velocities between the baryonic matter and dark matter may affect the details
of early structure formation; we do not include those effects here, although
this issue remains an important one to address in future simulations.

\begin{figure}
  \begin{centering}
  \includegraphics[width=\columnwidth]{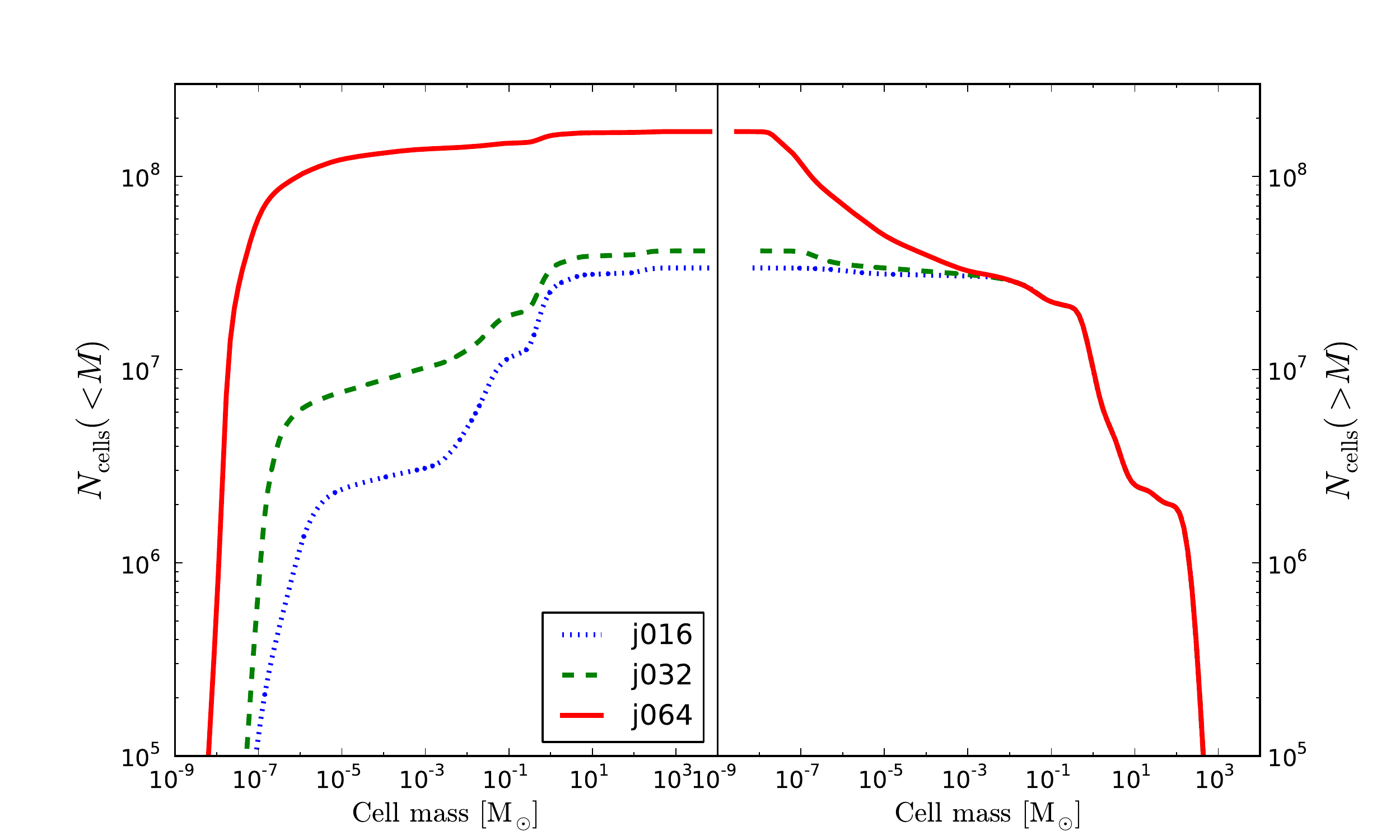}
  \caption{A plot of the number of cells as a function of mass enclosed in
  those cells, for each of the simulations at the time at which the simulations
  were terminated, at a peak density of $10^{-10}~\mathrm{g}~\mathrm{cm}^{-3}$,
  the time at which all analysis in this paper was conducted.  J16 is
  represented in dotted blue, J32 in dashed green, J64 in solid red.  This cell
  count is cumulative; on the left we show the total number of cells with mass
  less than the x-axis value, and on the right we show all cells with mass
  greater than the x-axis value.
  \label{fig:cell_count}}
  \end{centering}
\end{figure}

We have conducted four simulations, each of which vary only in the parameters
governing the AMR algorithm.  Typically in AMR calculations of the first stars
in the universe, Jeans resolution is the dominant refinement criteria.  The
four simulations presented in this work vary in the number of cells mandated to
resolve the Jeans Length in each direction; we have conducted simulations
requiring 16, 32, 64 and 128 cells in each direction, named J16, J32, J64 and
J128.  In all simulations, rather than calculate the Jeans length from the
local temperature, we calculated it assuming that the gas would be allowed to
cool to $200~\mathrm{K}$, as in \cite{2010ApJ...725L.140T}, which provides
substantially better resolution of higher-temperature gas.  We additionally
require refinement on relative overdensities; at the coarsest level, we flag
zones that are four times the mean density, and with each additional level this
decreases by a factor of $2^{-0.3}$.  This provides the refinement at scales
larger than the virial radius, inside which refinement based on Jeans criteria
dominates.   These simulations began from the same initial conditions and were
seeded with an identical uniform magnetic field of
$10^{-14}~\mathrm{G}~\mathrm{(proper)}$, corresponding to a comoving magnetic
field of $10^{-18}~\mathrm{G}$ at the start of our calculation.  With the
exception of J128, these simulations were terminated at a peak density of
$10^{-10}~\mathrm{g}~\mathrm{cm}^{-3}$; in the case of J128, we terminated the
simulation at a peak density of $10^{-19}~\mathrm{g}~\mathrm{cm}^{-3}$.  We
chose this density as it was the point at which the simulation ceased to be
feasible on available computational resources, and at which we were able to
demonstrate the variation in magnetic field amplification discussed below;
future studies will utilize this level of resolution, and potentially higher,
for calculations of Population III star formation.  In
Figure~\ref{fig:cell_count} we plot the number of cells in each simulation as a
function of the mass enclosed in that cell; for J64, we note that we have
$1.4\times10^{8}$ cells containing mass less than $10$ times the mass of the
Earth.

\section{Resolution Effects in Gas Dynamics}

The gas dynamics of these collapsing clouds is substantially modified by both
increased resolution and the indirect consequences of those
resolution-dependent effects.  By directly comparing the three mature
calculations (J16, J32, J64) at identical peak-density times, we quantify these
differences.  In particular, the differences in the chemical, kinetic, and
thermal states of the gas are the most pressing, as they will directly relate
to the potential for fragmentation at later times.  Because these simulations
are stopped before the formation of the first hydrostatic core, and more
importantly before that hydrostatic core has undergone several mass doublings,
we do not use these data sets to estimate the initial mass function of the
first stars, and instead confine our comments to measurable environmental
effects and their potential consequences for magnetic field amplification,
gravitational instability, and the subsequent accretion onto the central
molecular cloud.

\subsection{Resolution dependence of T and $\mathrm{H}_2$ and
$v_r$}\label{sec:resdep}

In the collapse of Population III stars, the chemical and thermal states of
the gas have been shown to be important when considering the accretion rates
and potential fragmentation \citep{ABN02, oshea07a, 2011Sci...331.1040C,
2011ApJ...737...75G, 2010ApJ...725L.140T, 2009Sci...325..601T}.  In particular,
recent findings from \cite{2011ApJ...726...55T} suggest that the character and
quantity of fragmentation in metal-free collapsing halos may depend strongly on
the behavior of molecular hydrogen at high densities.

As molecular hydrogen forms via three-body reactions (where typically the third
body is either another atom of hydrogen or a molecule of hydrogen) the
$4.48~\mathrm{eV}$ binding energy is released into the surrounding gas in
thermal energy, via collisional deexcitation.  During the collisional
dissociation of molecular hydrogen through the inverse reactions to the
three-body rates, the binding energy is removed from the thermal energy of the
gas.  \cite{2011Sci...331.1040C} proposes this as a mechanism for retaining
isothermality in collapsing halos; as primordial gas reaches densities above
$10^{-8}~\gcc$, it becomes optically thick to all efficient radiative cooling
until the gas reaches at least $6000~\mathrm{K}$.  This is proposed as a
mechanism for allowing the gas to fragment under gravitational instability.
This method of fragmentation relies heavily on the chemical state of the
molecular cloud at high densities, and would be sensitive to a lower molecular
hydrogen fraction, as this would impede gravitational instabilities from
retaining isothermality and thus serving as a channel for spontaneous
fragmentation.

\begin{figure*}
  \begin{centering}
  \includegraphics[width=0.54\textwidth]{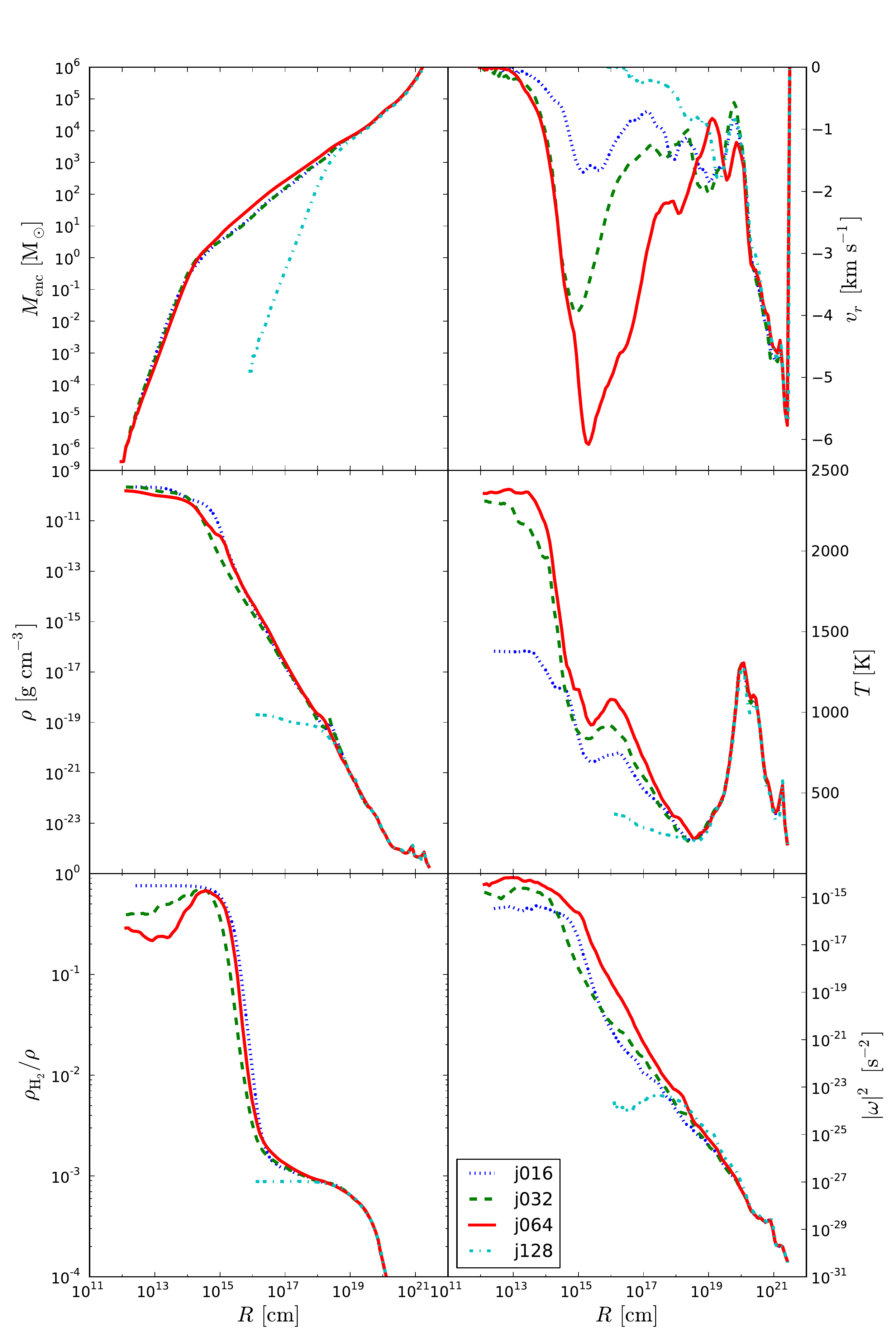}
  \caption{Radially-binned, spherically-averaged radial profiles of the four
  simulations taken at their final output.  These profiles were centered at the
  most dense point; J16 is dotted blue, J32 is dashed green, J64 is solid red,
  and J128 is dot dashed cyan.  Note that we terminated J128 before it reached
  the same peak density as the other simulations, although J16, J32 and J64
  were terminated at roughly identical peak densities.  Upper left is the total
  mass enclosed (in solar masses) at a given radius, upper right is the radial
  velocity in $\mathrm{km}~\mathrm{s}^{-1}$, middle left is the density inside
  a given radial bin in $\mathrm{g}~\mathrm{cm}^{-3}$, middle right is the
  temperature in K, bottom left is molecular hydrogen mass fraction, and bottom
  right is the magnitude of the local vorticity.
  \label{fig:final_output_radialprofs}}
  \end{centering}
\end{figure*}

In Figure~\ref{fig:final_output_radialprofs} we show spherically-averaged
radial profiles, computed at the final output and centered at the densest zone
in the calculation, of density, temperature, radial velocity, molecular
hydrogen mass fraction, vorticity and enclosed mass as a function of
radius.  The central velocity was subtracted before calculation of the radial
velocity.  As a function of radius, we see good agreement between J16, J32 and
J64 at radii greater than $10^{19}~\mathrm{cm}$ in the density profile.
However, at approximately $3\times10^{18}~\mathrm{cm}$ ($\sim1~\mathrm{pc}$) we
see a small bump in the J16 and J32 runs.  This is not present in the J64 runs.
Typically bumps in the density plot indicate the formation of a second clump or
other moderate level of fragmentation; however, based on the length and density
scales here, we attribute this to poor resolution in the lower-resolution runs
contributing to a spike of unresolved clumping just inside the virial scale; we
also note that it correlates directly with the lowest infall velocity in all
three simulations, so it may also simply be material that is being processed
into the inner regions of the halo.

The changes in the chemical and kinetic states of the gas as a function of
resolution are clearly correlated; with increased resolution, the infall
velocity increases, as does the corresponding temperature, causing a direct
decrease in the overall molecular hydrogen fraction.  While we still see a peak
in the molecular hydrogen, corresponding to a molecular hydrogen cloud of about
$1-5~\Msun$, the J64 run has substantially lower molecular hydrogen fraction of
$\sim0.2-0.3$ in the innermost core, versus nearly twice that for J32 and
approximately unity in the J16 case.  As is to be expected, this correlates
strongly with the thermal state of the gas; while J16 has a peak temperature of
approximately 1400K, the J32 and J64 cases both exceed 2000K (with J64 reaching
2400K.)  The rate at which molecular hydrogen is collisionally dissociated
steeply depends on temperature, and the relationship between these two
quantities has been explored previously \cite{2011ApJ...726...55T}.  With
increased resolution the homogeneity of the chemical and density structure of
the cloud cloud decreases, but we see more morphological homogeneity, as
discussed below.

The radial velocity structure remains roughly the same between all three
simulations, although the magnitude of the infall velocity increases with
increased resolution.  In the J16 case, we see an infall velocity at the radius
of the edge of the molecular cloud of approximately $2~\mathrm{km}/\mathrm{s}$,
although this increases to $6~\mathrm{km}/\mathrm{s}$ in the J64 case.  The J32
case is in between these two.  The J16 case shows slight peaking, indicating an
under-resolved shock at the exterior of the molecular region, but both J32 and
J64 show very strong shocking structures.  When compared to the enclosed mass,
we see that J64 encloses a factor of $\sim1.5$ as much mass at a given radius
inside the virial radius.  This likely contributes to this much greater buildup
in shock.  Furthermore, the increased resolution will allow much finer shocking
surfaces to develop and to be resolved (a noted problem with under-resolved
hydrodynamics.)

\subsection{Timescales of Collapse}

Ideally, in the course of a resolution study the basis for comparison between
two otherwise identical calculations would be comparison at a fixed time.
However, this method is not possible in full cosmological simulations because
the hydrodynamics of the calculation are greatly influenced by resolution
effects both directly, through turbulent stirring of the gas, viscosity, and
numerical diffusion effects, and indirectly through variations in the chemical
state of the gas as a result of temperature and turbulence differences.

In lieu of direct comparisons against identical times since the Big Bang, we
have conducted our analysis against identical peak density times.  While this 
neglects time dependencies such as the growth of mass at the innermost shocking
region, the velocity structure variations with time in the molecular cloud and
so on, it does provide a baseline for comparing the chemical and kinetic state
within the free-falling core.  In particular, this is the same technique
utilized in \cite{2011ApJ...726...55T}.

\begin{figure}
  \begin{centering}
  \includegraphics[width=0.9\columnwidth]{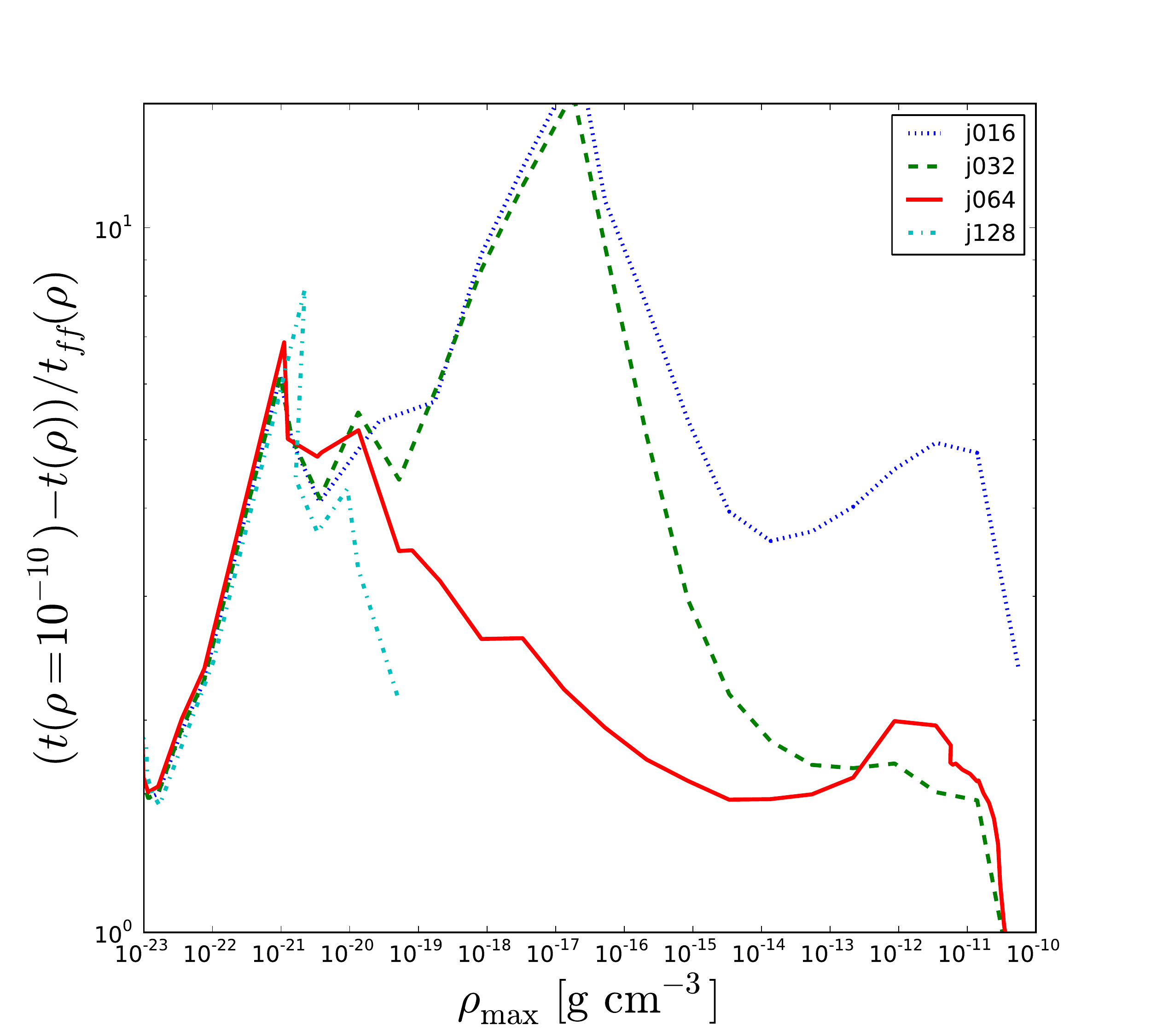}
  \caption{The differential free-fall time for all four simulations, J16
  (dotted blue), J32 (dashed green), J64 (solid red) and J128 (dashed-dotted
  cyan, which has not reached the same peak density)  The x-axis corresponds to
  a given density; the values along the y-axis have been calculated by
  subtracting the time at which the simulation was at a given peak density from
  the time at which the simulation had a peak density of of
  $10^{-10}~\mathrm{g}~\mathrm{cm}^{-3}$, then dividing by the differential
  free-fall time scales between the two times.  Higher-values therefore
  correspond to a longer collapse time, and lower-values correspond to a
  shorter collapse time with respect to the free-fall time.
  \label{fig:freefall}}
  \end{centering}
\end{figure}

To account for variations, we present in Figure~\ref{fig:freefall} a
visualization of the variations in the timescale of collapse and the freefall
timescale for each calculation.  Higher values correspond to a differential
collapse time compared to expected freefall timescales; one can clearly see
that simulations J16 and J32 are much slower to collapse than J64 at all
densities less than $10^{-13}~\mathrm{g}~\mathrm{cm}^{-3}$, at some densities
even reaching ten times the freefall timescale.  In particular, the peak delay
for simulations J16 and J32 occurs at a density of roughly
$10^{-17}~\mathrm{g}~\mathrm{cm}^{-3}$, the density at which (as discussed
above in \S~\ref{sec:resdep}) molecular hydrogen begins forming via the
three-body reaction, and therefore marking the edge of the molecular cloud.
Furthermore, as noted below, the morphological differences between the three
simulations indicate that this is correlated with variations in morphology.
 
\subsection{Morphological Differences}\label{sect:morphology}

\begin{figure*}
  \includegraphics[width=\textwidth]{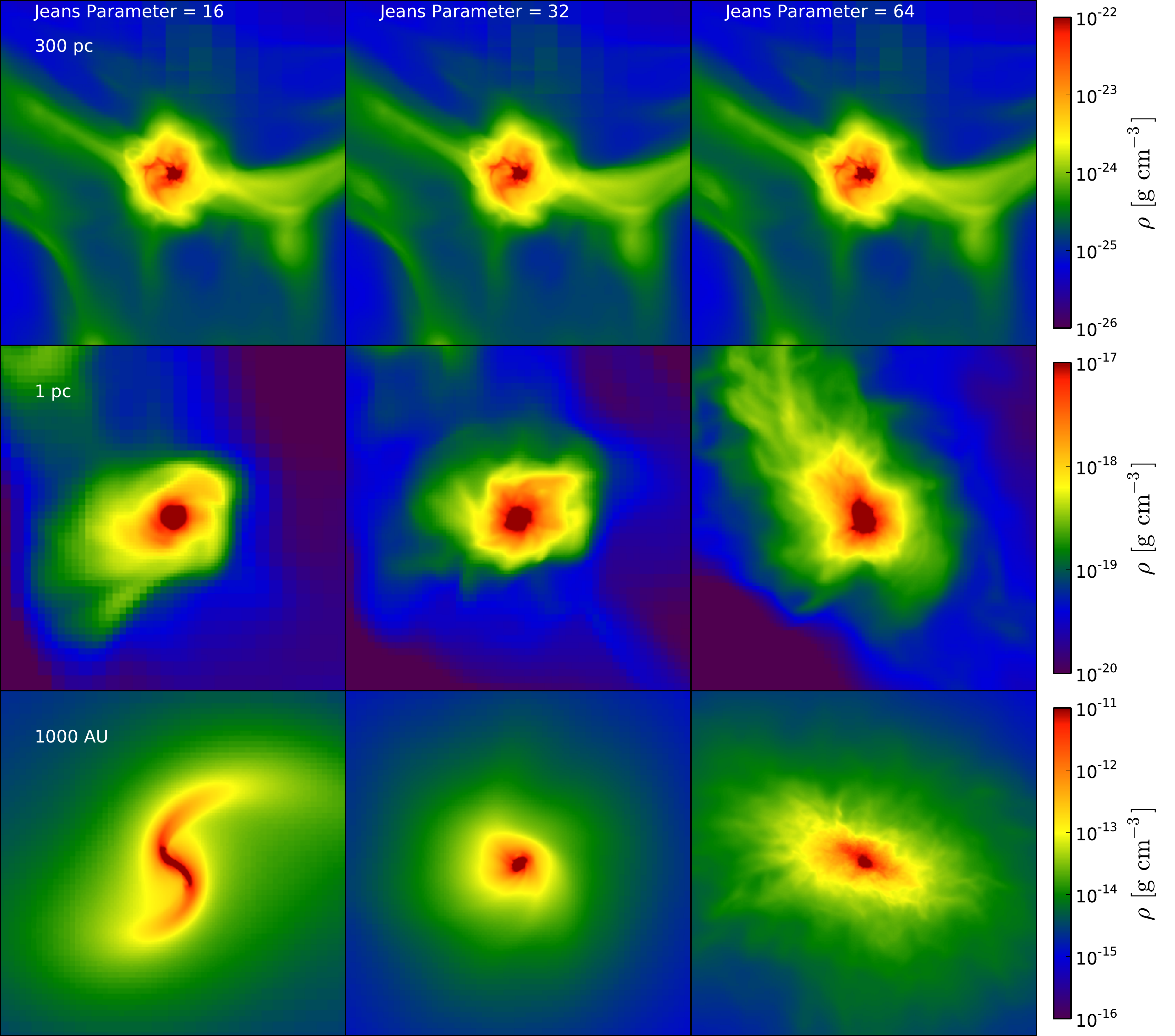}
  \caption{
  Density-weighted projections (through the entire simulation domain) of the
  average density field in simulations J16 (left column), J32 (second column)
  and J64 (third column) at fields of view of $300~\mathrm{pc}$ (top row),
  $1~\mathrm{pc}$ (middle row), and $1000~\mathrm{AU}$ (bottom row).
  \label{fig:rho_proj}}
\end{figure*}

In Figure~\ref{fig:rho_proj} we show density-weighted density projections of
the final output from the calculations.  From left, these show the J16,
J32 and J64 calculations, and from the top they have field of view of
$300~\mathrm{pc}$, $1~\mathrm{pc}$, and $1000~\mathrm{AU}$.  The most striking
difference between the simulations is found in the morphology.  Only one (J16)
of the three simulations that reached $10^{-9}~\mathrm{g}~\mathrm{cm}^{-3}$
(J16, J32, J64) has formed a clear disk, with two spiral arms.  Both J32 and
J64 have not yet formed a disk, and both show obvious cloud-like morphology,
with J64 exhibiting somewhat ellipsoidal morphology.  While J64 will form an
accretion disk at a later time, the variation in the settling times is
striking, particularly as the time at which an initial protostar ignites and
irradiates its surroundings will alter the environment for the potential
formation of subsequent protostars.  Furthermore, the dramatic differences in
the morphology suggest that resolution directly impacts the formation
environment, and may influence the gravitational stability of a collapsing
cloud.

\begin{figure*}
  \includegraphics[width=\textwidth]{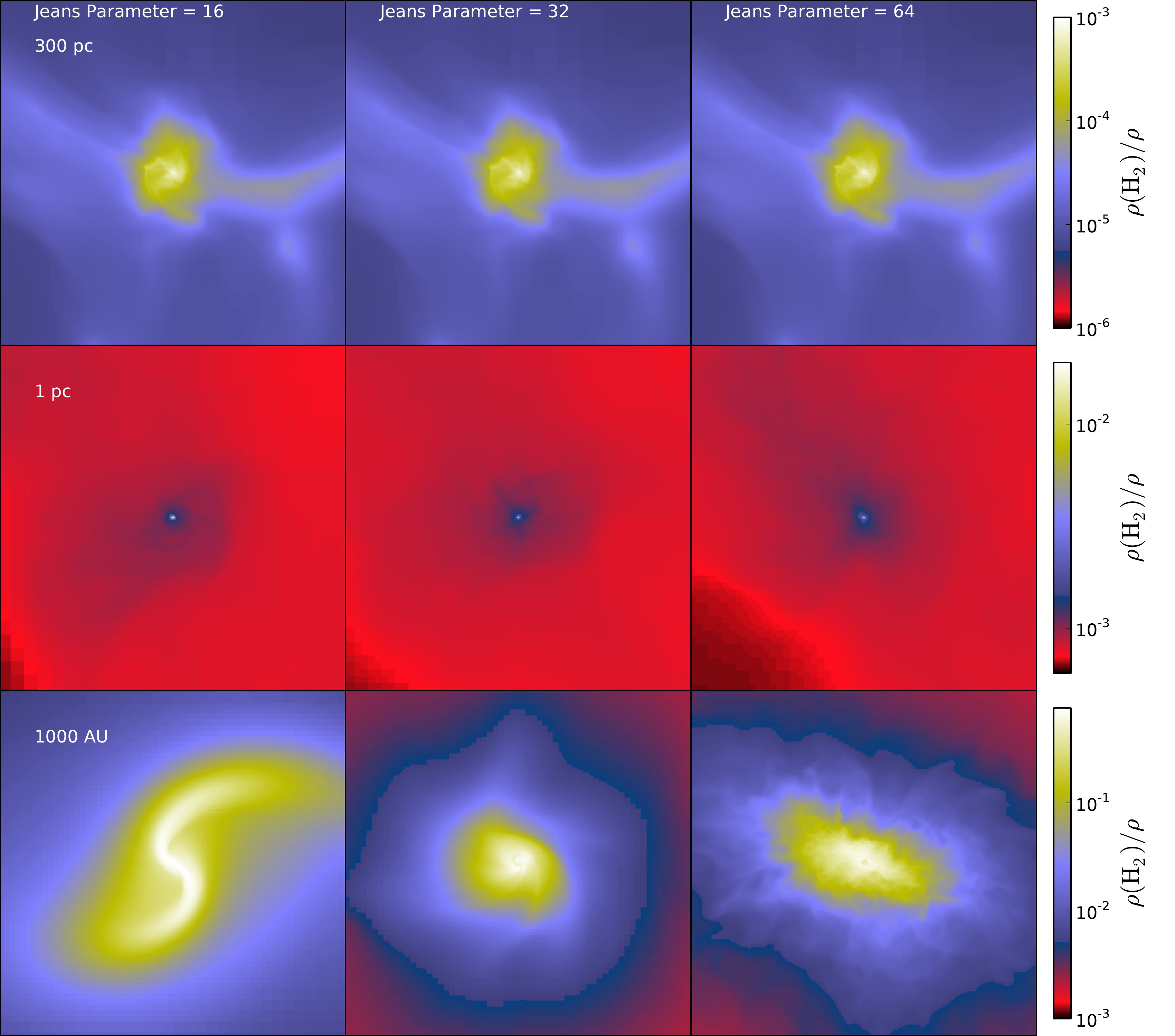}
  \caption{Density-weighted projections (through the entire simulation domain)
  of the average molecular hydrogen mass fraction field in simulations J16
  (left column), J32 (second column) and J64 (third column) column) at fields
  of view of $300~\mathrm{pc}$ (top row), $1~\mathrm{pc}$ (middle row), and
  $1000~\mathrm{AU}$ (bottom row).
  \label{fig:h2f_proj}}
\end{figure*}

In Figure~\ref{fig:h2f_proj} we plot density-weighted projections of the
molecular hydrogen mass fraction from the same outputs and with the same field
of views as in Figure~\ref{fig:rho_proj}.  We note in particular that the
morphology of the molecular hydrogen cloud roughly tracks the overall density
structure; however, the molecular cloud in the J16 simulation is substantially
larger than those in the J32 and J64 calculations.  This can be seen in
Figure~\ref{fig:final_output_radialprofs}, wherein the J16 calculation has a
slightly larger molecular cloud when averaged radially.  Furthermore, while not
obvious here owing to the density-weighting, the innermost regions of J32 and
J64 show molecular hydrogen fractions of $\leq0.4$, and even as low as $0.2$ in
J64.  In J16, the molecular hydrogen cloud tracks almost identically the spiral
arms in density; at the leading and trailing ends the fraction drops.  This
occurs because, as the three-body reaction for forming molecular hydrogen is
strongly density-dependent.  The increased cooling properties with higher
molecular hydrogen fraction will only exacerbate this distinction between the
spiral arms and the medium of the disk, spurring further collapse.  In the J32
and J64 runs, we see not only highly-irregular, spheroidal structure, but we
see little to no evidence for runaway gravitational instability at this time in
the collapse; in fact, the roughly spheroidal structure of the cloud suggests
that disk fragmentation at the $\sim100~\mathrm{AU}$ scale may be disfavored,
certainly until a later time.

\section{Magnetic Field Amplification}

\begin{figure}
\begin{centering}
  \includegraphics[width=\columnwidth]{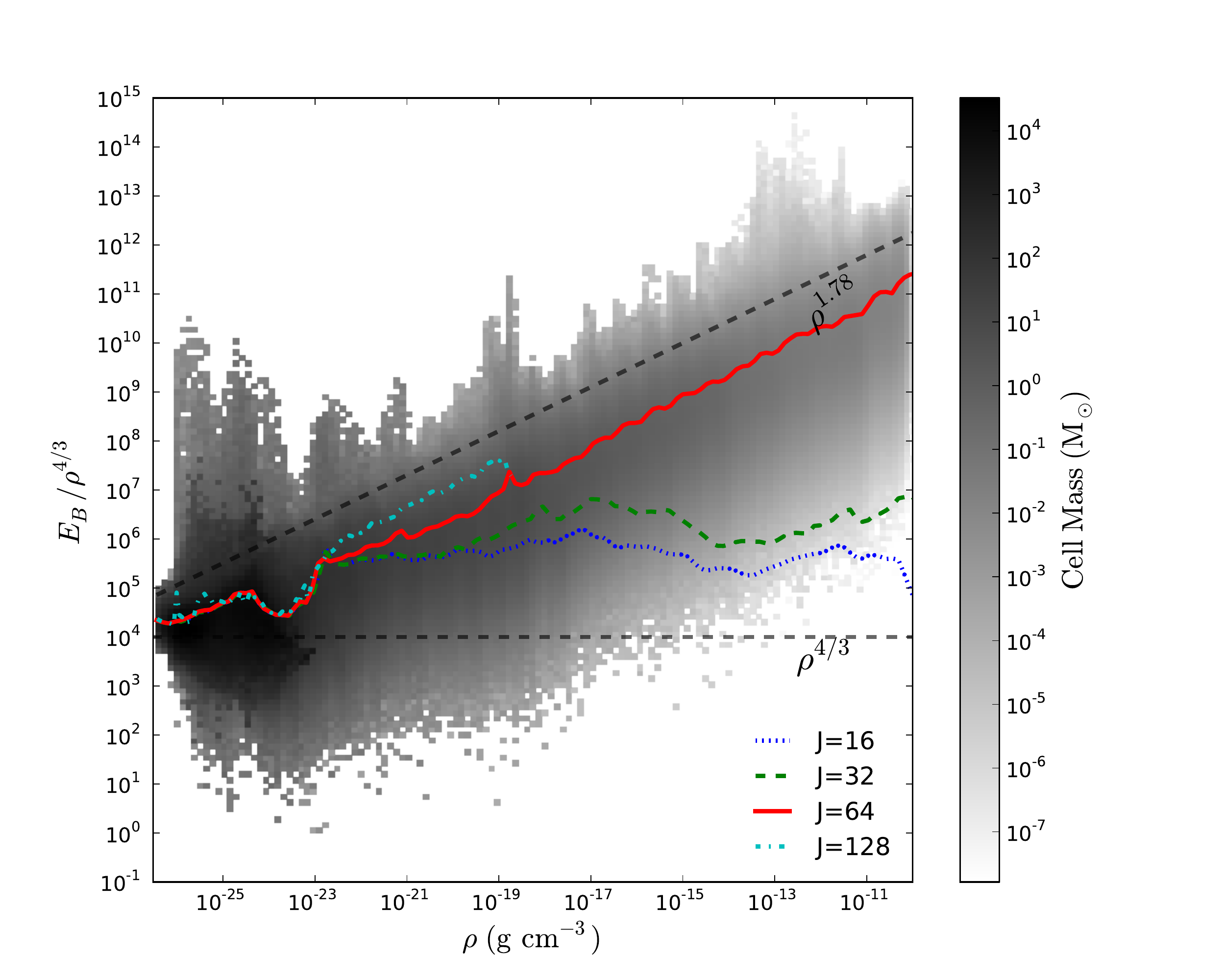}
  \caption{Lines show volume-weighted magnetic energy scaled by $\rho^{4/3}$ as a function of density
    for a $1-\mathrm{kpc}$ sphere centered on the densest point in the final
    output for each of four resolutions, J16, J32, J64, and J128. The colored
    pixels give the mass in $E_B - \rho$ bins for the J64 run. We have overlaid
    two power laws, giving $B \propto \rho^{4/3}$ (lower dashed line), the
    expected scaling for spherical collapse of frozen-in magnetic fields, and $B
    \propto \rho^{1.78}$ (upper dashed line), the best-fit power law to the J64
    data.
    \label{fig:magen_radialprofs}
  }
\end{centering}
\end{figure}

We infer magnetic field amplification above that arising simply from the
spherical compression of frozen-in field lines during collapse by plotting the
magnetic energy, $E_B = B^2/8\pi$, as a function of density for each of our four simulations
(figure~\ref{fig:magen_radialprofs}). Assuming
a power law relation between magnetic energy and density, $E_B \propto
\rho^{b}$, a spherical collapse would result in $b = 4/3$. Any steeper values of
$b$ must result in additional amplification in the form of a dynamo. We expect a
small-scale turbulent dynamo if there is significant kinetic energy in turbulent
motions. Figure~\ref{fig:magen_radialprofs} shows that for the best resolved
simulations, J64 and J128, the magnetic field scales with a much higher power,
$b \simeq 1.78$ at all densities $\rho \gtrsim 10^{-26}$, roughly
consistent with the findings of \cite{2011ApJ...731...62F}. By contrast, the
lower resolution runs (J16 and J32) begin to show $b > 4/3$ behavior at the
same density as the higher resolution runs, but never reach the same $b$ as the
higher runs and eventually begin to show a shallower slope of $E_B$ with $\rho$
at the highest densities ($\rho \gtrsim 10^{-15}$).  In the outer part of the
collapse, near the virial radius, all runs show $b \simeq 4/3$, suggesting that
at low densities, the growth of magnetic energy is primarily due to the roughly
spherical collapse, though the data is somewhat scant.  

\subsection{Velocity Structure In the Collapse Region}

\begin{figure*}
  \includegraphics[width=\textwidth]{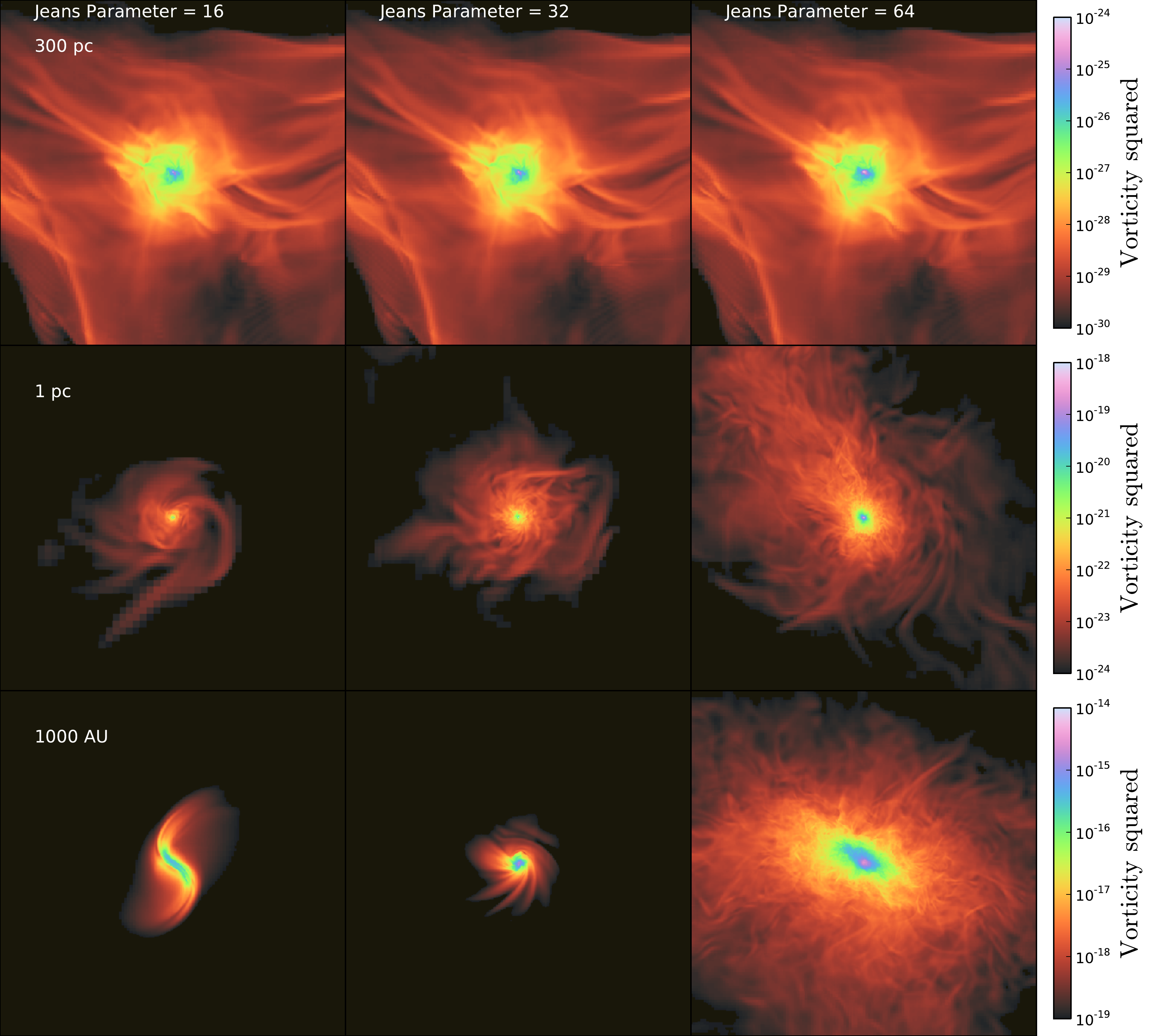}
  \caption{
  Density-weighted projections (through the entire simulation domain) of the
  average magnitude of the local vorticity in simulations J16 (left column),
  J32 (second column) and J64 (third column) at fields of view of
  $300~\mathrm{pc}$ (top row), $1~\mathrm{pc}$ (middle row), and
  $1000~\mathrm{AU}$ (bottom row).  \label{fig:vort_proj}}
\end{figure*}

The dynamo activity in the previous section is resolution dependent, but robust
above roughly 64 cells per Jeans length (J64); that is, we continue to see more
field amplification, consistent with a minimum resolution requirement of
between 32 and 64 zones per Jeans length.  This is broadly consonant with the
results of Federrath et al, who find a similar resolution cutoff in their
simulations of magnetic field growth in the nearly-isothermal collapse of a
Bonnor-Ebert sphere. In their picture, the reason for this cutoff is the lack
of sufficient power in rotational motions when the Jeans length is resolved
with fewer than $\sim 30$ cells. Our resolution requirement for strong and
sustained dynamo action appears to be about a factor of two larger, which could
be due to our use of a considerably more diffusive MHD solver (we use the HLL
solver; they use HLL3R). In order to solidify the connection between dynamo
action and the presence or absence of turbulent fluctuations, we first consider
projections of $\mathbf{\omega}^2$ where $\mathbf{\omega = \nabla \times v}$ is
the fluid vorticity centered at the densest point in the computation
(figure~\ref{fig:vort_proj}). At the $300 \mathrm{pc}$ scale
(figure~\ref{fig:vort_proj}, top row), the vorticity is identical among the
four resolutions. However, zooming in to $\sim 1 \mathrm{pc}$, there are
already significant differences between all four runs. For the other three, a
trend is clear: increasing resolution creates much larger regions of high
vorticity that is indicative both of an overall increase in the turbulent
energy but also of a decreasing coherence of the collapse region (see
section~\ref{sect:morphology}). Finally, at $1000 \mathrm{AU}$, the vorticity
structure is completely different between the ordered disk-like structure in
J16 and J32 and the amorphous turbulent core of J64. It is thus clear that as a
function of resolution, we see an increase in vorticity production and a
decrease in the characteristic scale of that vorticity at the smallest scales
within the collapsing core. These high density regions are just where magnetic
energy growth with density begin to fall off in low resolution simulations.
This correlation between vorticity and magnetic energy is consistent with
small-scale dynamo action generated by incoherent velocity fields.

\begin{figure*}
  \includegraphics[width=\textwidth]{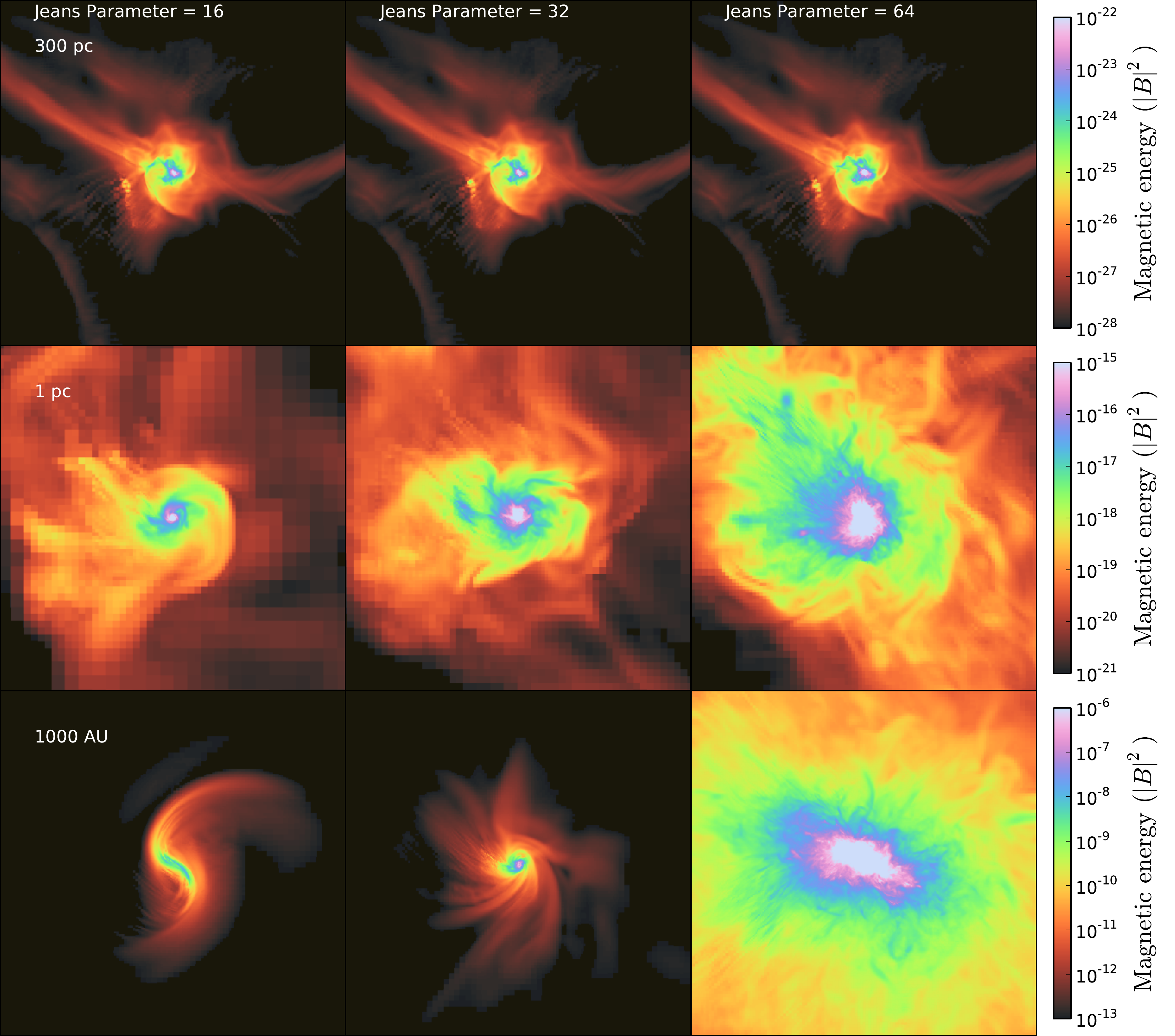}
  \caption{
  Density-weighted projections (through the entire simulation domain) of the
  average magnetic energy in simulations J16 (left column), J32 (second column)
  and J64 (third column) at fields of view of $300~\mathrm{pc}$ (top row),
  $1~\mathrm{pc}$ (middle row), and $1000~\mathrm{AU}$ (bottom row).
  \label{fig:magen_proj}}
\end{figure*}

\subsection{Magnetic Field Saturation}
The kinematic, small-scale dynamo acts by twisting magnetic field in a turbulent
flow field. The random-walk character of the flow will lead to a continuous
stretching and thus strengthening of the magnetic field. In a typical numerical
model of the small-scale dynamo, turbulence is driven in a fluid with a tiny
seed field; the field grows exponentially until it nears equipartition with the
turbulent velocity field, at which point the Lorentz force reacts back on the
fluid, leading to a strongly nonlinear coupling between $\mathbf{v}$ and
$\mathbf{B}$ which ultimately saturates the growth of the magnetic field. 

Here, the collapse timescale is decreasing with time, which means that the
computation reaches its end long before the magnetic field reaches
saturation. We stop each of our runs when their peak density is $\rho_{max}
\simeq 2.2 \times 10^{-10}$, except for the J128 run, which was terminated at
$\rho_{max} \simeq 2.0\times10^{-19}$ due to constraints on computing
time. Nevertheless, even in the J64 run, where dynamo action is most vigorous,
the kinetic energy exceeds the magnetic energy in mass weighted averages by a
factor of $\sim 3600$. 

The growth rate $\sigma$ of the small-scale dynamo is a function of the Reynolds
number $\mathrm{Re}$, the ratio of the advective timescale to the viscous
timescale, which we do not control in these runs (dissipation is solely a
numerical artifact of finite resolution). \citep[][see, e.g.,]{2011ApJ...731...62F}
found the growth rate to be $\sigma \propto \mathrm{Re}^{0.3}$. Even if we were
to solve the Navier-Stokes equation including viscosity and the non-ideal
induction equation including Ohmic resistivity, we would be nowhere near the
actual Reynolds numbers of the flows in primordial gas. A rough estimate of
$\mathrm{Re}$ using the Spitzer viscosity, characteristic velocity and length
scales taken from Figure~\ref{fig:final_output_radialprofs} is about
$\mathrm{Re \sim 10^{12}}$ at a radius of $\sim 1000\ \mathrm{AU}$ . This means
that our simulations cannot make a definitive statement regarding the
saturation level of the collapse dynamo, though our results can be taken as a
lower limit to the field strength during collapse.  We note that this indicates
that magnetic fields may be dynamically important in the formation of the first
stars, and that their amplification requires much higher resolution than is
currently being utilized.

\section{Discussion}

It is clear from our simulations that gravitational collapse from cosmological
initial conditions does indeed drive a small-scale dynamo if the flow is
resolved well enough to capture the turbulent fluctuations. This turbulence is
produced self-consistently by the collapse dynamics, and is not seeded by any
driving or initial perturbations.  The change in slope between J64 and J128
indicates that while we have demonstrated the minimum resolution for dynamo
action to occur, further amplification of the magnetic field will result from
increased resolution, and thus our results should not be considered converged.
We will explore the details of the generation the turbulence and the details of
the dynamo generated magnetic field in a forthcoming paper.

The most striking difference between the J16, J32 and J64 runs is the
resolution-dependence in the chemical and kinetic states of the gas.  While we
initially expected that the simulations would produce largely the same results
in the structure of the collapsing clouds, the three simulations produced
substantially different disk structure.  Because we examined these three
simulations at identical peak density values, the time remaining until the
formation of the first hydrostatic core should be roughly identical in each
run; as such, the initial structure of the cloud will likely not change in that
intervening time.  \cite{2011Sci...331.1040C} presented simple
comparisons between the accretion rate onto the central accretion disk and the
accretion rate onto a central protostar; in those calculations these
comparisons indicate that the central core will be unable to process sufficient
gas to deplete the disk, which they observed result in gravitational
instability.  We note that in our simulations, we see substantially reduced
disk-like structure, as well as substantially higher infall velocities at the
edge of the molecular hydrogen cloud.  The combination of these two effects
results in an unclear change to the potential fragmentation characteristics of
these collapsing clouds; reduced angular momentum should allow greater
accretion onto a central core, but the increased radial velocity onto the
molecular cloud will result in greater mass-buildup.  Furthermore, we note
that while we do not resolve the formation of a central core or the subsequent
accretion, and thus cannot make statements regarding the fragmentation
characteristics in our simulation, the result presented in
\cite{2011Sci...331.1040C} was constructed from a series of fixed resolution
calculations with a much lower Jeans resolution than in our J64 run (and indeed
our J32 run); the results in this paper suggest that higher resolution may
change the specific nature of fragmentation.

The correlation between the dissociation of molecular hydrogen and the
increased resolution was unexpected, but in retrospect is not entirely
surprising.  With increased infall velocity, increased turbulent support, and
increased overall temperature, the rate at which molecular hydrogen
collisionally dissociates increases steeply, while the rate at which it
associates via three-body reactions decreases.  With increased turbulent
energy, and a corresponding increase in the temperature, the innermost region
in the pre-stellar molecular cloud will naturally undergo some dissociation.
However, what remains unclear is how this will affect the response of the gas
to gravitational instability.  Prior to the work of \cite{2011Sci...331.1040C,
2011ApJ...737...75G}, fragmentation had been seen at scales of
$\sim2000~\mathrm{AU}$ \citep{2009Sci...325..601T, 2010MNRAS.403...45S}; at
smaller scales, fragmentation had (prior to the works of Clark, Greif and their
collaborators) been believed to be suppressed because of the inefficient
cooling of molecular hydrogen.  In these more recent works, however, the
ability of the gas to shed thermal energy through dissociation of molecular
hydrogen, essentially allowing compressional heating to be freely reversed,
has been identified as the mechanism by which fragmentation proceeds.  At high
densities (the densities at which sink particles have been inserted in their
calculations), the gas becomes effectively isothermal (and later, once the
molecular hydrogen has been depleted, adiabatic.)  However, this hypothesis
relies on a substantial reservoir of molecular hydrogen; while a gas cloud
containing a depleted but non-zero supply of molecular hydrogen is still likely
to undergo a quasi-isothermal phase during its collapse, the resilience of that
gas and the stiffness of its effective equation of state are likely to be
modified.  Furthermore, because the fraction of molecular hydrogen, and thus
the equation of state of potentially-fragmenting gas sensitively depends on the
rate at which molecular hydrogen is formed and destroyed, as discussed in
\cite{2011ApJ...726...55T}, this provides an additional uncertainty.  Without
extremely high-resolution calculations that follow the formation of the first
core, as well as its subsequent accretion over several thousand years, the
character of fragmentation in accretion disks around Population III pre-stellar
cores is still an open question.


\section{Conclusions}

We have presented on the first fully-cosmological calculations of Population
III star formation that include all relevant chemical processes, as well as
magnetic fields.  We see, in agreement with \cite{2011ApJ...731...62F}, that a
critical resolution exists above which we are able to resolve small-scale
dynamo action resulting in increased magnetic-field field growth.  While we
adequately resolve the \textit{action} of this small-scale dynamo, we have not
yet resolved the amplification caused by that dynamo, as demonstrated by the
nascent J128 simulation.  In fact, while these calculations have yet to show a
dynamically-important magnetic field, estimates of the possible saturation
level of the magnetic field indicate that with increased resolution or stronger
seed fields, magnetic fields may in fact become dynamically important.

A side effect of conducting these resolution studies has been that we see
substantial variation in the chemical, kinetic, and velocity structure of
collapsing metal-free, star-forming clouds.  In particular, the substantial
difference in the J64 run suggests that under-resolving the hydrodynamics can
result in incorrect solutions: the fraction of hydrogen gas that is in
molecules, the speed of sound, the infall velocity and the turbulent support
all depend strongly on hydrodynamic resolution.

Our calculations, while falling short of following potential fragmentation in
detail, suggest that resolution effects may be an additional complication in
determining the initial mass function of the first stars; in fact, they suggest
that previous calculations in the literature have not yet converged on a
solution.  Future calculations, of much higher resolution and attaining a
greater saturation level of magnetic fields, will help to resolve these
outstanding questions.

\acknowledgments

M.J.T.~acknowledges support by the NSF CI TraCS fellowship award OCI-1048505.
This material is based upon work supported by the National Science Foundation
under a grant awarded in 2010.
G.L.B.~acknowledges support from NSF grants AST-0547823, AST-0908390, and
AST-1008134, as well as computational resources from NASA, the NSF Teragrid,
and Columbia University’s Hotfoot cluster.  The simulations were run at the
SLAC National Accelerator Laboratory, on the computing cluster orange.  We are
grateful for the support of the scientific computing department at SLAC as well
as to Stuart Marshall's patient assistance with file system and cluster issues.
All analysis was conducted using \texttt{yt} (\cite{yt_full_paper},
\texttt{http://yt-project.org/}).  We thank the anonymous referee for useful
comments.


\label{lastpage}

\end{document}